\documentclass[aps,twocolumn,10pt]{revtex4-2}
\usepackage[utf8]{inputenc}
\usepackage[svgnames]{xcolor}
\usepackage[most]{tcolorbox}
\usepackage{caption}
\tcbset{colback=PaleGreen!0!white, colframe=NavyBlue, 
	highlight math style= {enhanced, 
		colframe=red,colback=red!10!white,boxsep=0pt}
}
\usepackage{relsize}

\usepackage{amsmath}
\usepackage[bottom]{footmisc}
\usepackage{braket}
\usepackage{graphicx}
\usepackage{mwe}
\usepackage[section]{placeins}
\usepackage{framed}
\usepackage{csquotes}
\usepackage{tikz}
\usetikzlibrary{decorations.markings}
\usetikzlibrary{decorations.pathmorphing}
\definecolor{shadecolor}{rgb}{0.90,0.90,0.90}
\usepackage{hyperref}
\usepackage{bbold}
\usepackage{subcaption}
\usepackage{pifont}
\usepackage{setspace}
\usepackage{amsmath, amssymb, amsthm, float, graphicx,amsfonts}
\usepackage{amsthm}
\usepackage{array, boldline, makecell, booktabs}

\theoremstyle{definition}

\hypersetup{colorlinks=true, linkcolor=DarkRed, citecolor=DarkRed,urlcolor=Indigo,linktocpage}
\def\beq{\begin{eqnarray}}\def\eeq{\end{eqnarray}}
\def\be{\begin{equation}}\def\ee{\end{equation}}
\def\bs{\begin{split}}\def\es{\end{split}}

\def\a{\alpha}
\def\e{\epsilon}

\usepackage{bm}

\usepackage{lipsum, babel}

\usepackage{amsthm}
\usepackage{array, boldline, makecell, booktabs}
\usepackage{amstext} 
\usepackage{array}   
\newcolumntype{C}{>{$}c<{$}}
\newcolumntype{L}{>{$}l<{$}} 

\begin{document}

\title{\bf Bootstrapping leading hadronic muon anomaly}
\author{Ahmadullah Zahed\\
\it ICTP, International Centre for Theoretical Physics,\\
\it Strada Costiera 11, 34135, Trieste, Italy.\\
azahed@ictp.it }

\begin{abstract}{
We bootstrap the leading order hadronic contribution to $a_\mu$ using unitarity, analytic properties, crossing symmetry and finite energy sum rules (FESR) from quantum chromodynamics (QCD), establishing a lower bound. Combining this lower bound with the remaining precisely calculated contributions from quantum electrodynamics and electroweak interactions, we achieve a lower bound on muon anomaly $a_\mu$. {Since the FESRs have uncertainties, our bound depends on the choices of FESRs within these uncertainties. A conservative choice of the FESR gives a conservative lower bound, consistent with Standard Model (SM) data-driven prediction. We show that there are other valid choices of FESRs within the uncertainties that lead to lower bounds, which are inconsistent with SM data-driven prediction but consistent with the measured values of the muon anomaly.} {The bootstrapped spectral density shows a $\rho$-resonance peak similar to experimental hadronic cross-ratio data, providing a bootstrap prediction for $\rho$-meson mass.}
}
\end{abstract}
\maketitle

\section{Introduction}
The muon anomaly $a_\mu=(g-2)_\mu/2$ encapsulates how the muon interacts with magnetic fields through its intrinsic spin. The measurements of the muon anomaly \cite{exp1,exp2} show a deviation from the theoretical prediction up to $5.0\sigma$ \cite{th1} while agreeing with the lattice QCD simulations within $0.9\sigma$ \cite{lat1,lat2}. 

A significant contribution to this discrepancy arises from the hadronic vacuum polarisation (HVP) at the leading order in the fine-structure constant ($a_\mu^{\text{LO-HVP}}$), where the muon's interaction is influenced by the complex interplay of quarks and gluons through the strong force, as described by quantum chromodynamics (QCD). The hadronic contribution is more elusive due to QCD's strongly coupled, non-perturbative nature at low energies, unlike the electromagnetic and electroweak contributions, which can be calculated with great precision. This makes the precise evaluation of the hadronic effects a central challenge to understand  $a_\mu$  and its implications for particle physics. 

The bootstrap approach in quantum field theory (QFT) is a non-perturbative framework utilizing basic principles like unitarity, analyticity, crossing and other symmetries of QFT to constrain a theory space \cite{boot_white, cftboot_white}. We provide a bootstrap approach to the hadronic contribution by imposing the unitary condition among spectral density, pion partial wave and form factor as a positive semi-definite condition \cite{joao_semi,Heboth} while incorporating finite energy sum rules (FESR) from QCD.
\begin{figure}[hbt!]
     \centering
         \includegraphics[scale=0.23]{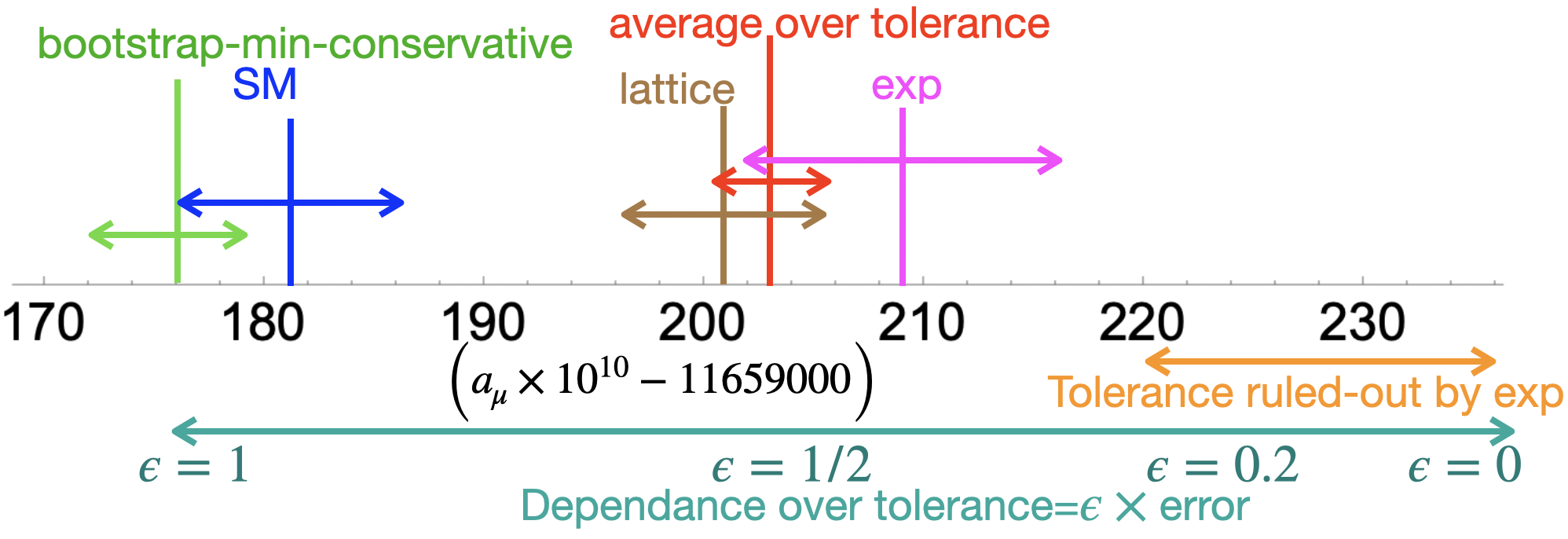}
        \caption{Comparison: The SM prediction (dark blue) within the error bars saturates our conservative lower bound (green). The dependence on the tolerance is shown in light blue, $\e$ from 1 to 0; almost all the choices ( $0.9$ to $0$ ) are incompatible with SM predictions, except for the weakest ones. Nearly all tolerance choices (1 to $0.2$) are consistent with the measured value (magenta); tolerance ruled out by the exp is shown in orange ($0.2$ to $0$). The average bound (red) is incompatible with the SM prediction and is saturated by the lattice result (brown) and within the error bars of the measured value (magenta).}
        \label{fig:bootmuong}
\end{figure}

The FESR have errors due to the QCD parameters, mainly from the dimension-four gluon condensate and dimension-six quark condensates. Hence, we input the FESRs as inequality up to a tolerance. A conservative choice of the tolerance is the error itself (namely $mean-error<FESR<mean+error$), resulting in a lower bound that aligns with the dispersive-data-driven standard model (SM) prediction within the error bars. The conservative lower bound is
$$
a_\mu^\text{bootstrap-min-conservative}=11659176.3^{+3}_{-3}\times 10^{-10}\,.
$$ There are two distinct sources of uncertainty in our determination of the lower bound: 1. Numerical/methodological uncertainties from the bootstrap implementation, and 2. Uncertainty from the input FESRs. Here, the uncertainty presented in the lower bound is due to the methodological errors, since the uncertainty from the FESRs has already been taken care of by imposing them as inequalities. We obtain a lower bound on the muon anomaly and a corresponding spectral density that satisfies essential physical constraints-namely unitarity, analyticity, crossing symmetry, and the FESRs (within uncertainties)-while achieving this minimum. Remarkably, the bootstrapped spectral density exhibits a $\rho$-resonance-like peak similar to that seen in the experimental hadronic cross-section data, allowing a prediction of the $\rho$-meson mass. \textit{This is an impressive bootstrap result since we do not provide any input for the $\rho$-resonance.} 

Since a conservative choice of FESR gives a lower bound saturated by the SM data-driven value, we are naturally led to ask whether other choice of the tolerance leads to a lower bound that is inconsistent with the SM data-driven value but consistent with the measured value. Since randomly scanning over the FESRs within uncertainties leads to lower bounds that are inconsistent with the SM data-driven value and sometimes even measured value, a more systematic way would be to move away from the conservative choice in small steps by scanning $\epsilon$, namely writing $mean-\epsilon\times error<FESR<mean+\epsilon\times error$. Figure \eqref{fig:bootmuong} shows the dependence on tolerance$=\e\times$error by varying $\e$ from 1 to 0; almost all the choices are incompatible with the SM predictions \cite{th1}, except for the weakest ones--see also table \eqref{tab:e}. Note that after the minimisation process, the optimal solution should return a value within the interval $[ mean - \epsilon\times error,~ mean +\epsilon\times error]$. Scanning $\epsilon$ from 1 to 0 systematically explores a range of possible FESR values within these uncertainties. So, dependence on the tolerance (presented in figure \eqref{fig:bootmuong})  is a conservative way of presenting the uncertainty of the lower bound coming from FESRs. The improvement due to the tolerance is evident because the mentioned QCD condensates are poorly determined and lack first-principle computations. Average over tolerance appears around error/2,  gives $$a_\mu^\text{min-average}=11659204.3^{+1.6}_{-1.6}\times 10^{-10}\,,$$which is incompatible with the SM prediction, while saturated by both the lattice computation and the measured value within the error bars. The average over tolerance is introduced as a heuristic benchmark: it provides a representative value that smooths out the dependence from individual $\epsilon$ values and gives a sense of the ``central tendency" of the lower bounds. While this average does not have a rigorous statistical interpretation, it helps to illustrate the overall trend. Figure \eqref{fig:bootmuong} summarises our findings.

From figure \eqref{fig:bootmuong}, it is evident that some choices of the tolerances are ruled out by the measured values of the muon anomaly; roughly, tolerances smaller than $\epsilon=0.2$ are ruled out. This eventually rules out some values of the QCD condensates (mainly dimension-four gluon condensate and dimension-six quark condensates, which are poorly determined due to a lack of first-principle computations). We quote corresponding ruled-out values in the sections below after these have been defined--see section III and figure \eqref{fig:condensates}.

\section{Bootstrapping leading hadronic contribution to muon anomaly}
The leading hadronic contribution to the muon anomaly is given by
\begin{equation}
\label{def:a_muhad}
    a_\mu^\text{LO-HVP} = \frac{4\alpha^2}{\pi} \int_{4m_\pi^2}^{\infty}\frac{K(t) \text{Im}\Pi(t)}{t} \, dt\,,
\end{equation}
where  $\Pi(t)$ is the hadronic vacuum polarisation (HVP) and $K(t)=\int_{0}^{1}dx \frac{x^2(1-x)}{x^2+(1-x)t/m^2_\mu}$.  We set energy unit such that $m_\pi=1$ for convenience. 

The unitary condition among the Im$\Pi(t)$, pion partial wave and form factor is given by\cite{joao_semi,Heboth}
\be\label{semimat}
B(s)\equiv\left(
\begin{array}{ccc}
1 & S_1^1(s) & \mathcal{F}_1^1(s) \\
S_1^{1*}(s) & 1 & \mathcal{F}_1^{1*}(s) \\
\mathcal{F}_1^{1*}(s) & \mathcal{F}_1^{1}(s) & \rho_1^1(s)
\end{array}
\right) \succeq 0, \quad s > 4\,,
\ee
where $\rho^1_1(s)\times \frac{(2\pi)^4}{s}=\text{Im}\Pi(s)$ and $\mathcal{F}_1^1(s)=\frac{\sqrt{\frac{4 \pi }{3}} \left(\frac{s-4}{4}\right)^{3/4}}{\left(8 \pi ^3\right) \sqrt[4]{s}}F(s) $, 
with $F(s)$ being some vector form factor normalized as $F(0)=1$. 
The P-wave $S_1^1(s)$ is non-trivially related to other isospin ($I$) and spin ($\ell$) partial waves through the unitary relation $|S_\ell^I(s)|\leq 1$. The analyticity and crossing symmetry of pion scattering amplitudes are used to compute the pion partial waves \cite{qcdboot1,sboot3}--see also\cite{aninda1,mehmet1,me1,me2}. The unitary condition \eqref{semimat} is a generalization of Watson's equation\cite{Watson:1954uc} and was introduced in \cite{joao_semi}, further developed and introduced QCD constraints for pion bootstrap in \cite{Heboth} which plays an important role in our analysis. For derivation and details, we refer to \cite{Heboth} keeping in mind  $J_\mu=\sum_{q=u,d,s} e_q \bar{q}\gamma_\mu q$. We use the finite energy sum rules (FESRs) from the QCD constraints. The FESRs for each quark contribution \cite{SVZ, SVZ2} add up with appropriate pre-factors to provide the FESRs for  $\int_{4}^{s_0}t^{n}\text{Im}\Pi(t) dt$ \cite{qcdpaper} for $n=0,1,2$-- see appendix \ref{ap:qcdsumrules}. The choice of the $s_0$ is crucial. The lower $s_0$ value gives a better lower bound\cite{qcdpaper}. However, we can't go arbitrarily low in $s_0$. Below $s_0=1.19$ GeV${}^2$, the strange quark FESRs start violating simple positivity inequality derived from Holder's inequality \cite{qcdpaper}. Hence, we stop at $s_0=1.19$ GeV${}^2$.

\textcolor{Blue}{{Bootstrap strategy:}} \textit{Utilize the unitary condition \eqref{semimat}, analytic properties, sum rules for  $\int_{4}^{s_0}t^{n} ~\mathrm{Im}\Pi(s) dt$ and the partial wave unitarity $|S_\ell^I(s)|\leq 1$ to scan the space of $S_\ell^{I}(s), ~\mathrm{Im}\Pi(s)$ and $ \mathcal{F}^1_1(s)$ which minimizes the $a_\mu^\text{LO-HVP}$.} The scanning of space of $S_\ell^{}(s), ~\text{Im}\Pi(s)=\rho^1_1(s)\times \frac{(2\pi)^4}{s}$ and $ \mathcal{F}^1_1(s)$ is done by writing down a suitable ansatz as a sum over a basis. The crossing symmetry and analyticity dictates basis for $S_\ell^{I}(s)$ \cite{qcdboot1}, while analyticity dictates basis for $\rho_1^1(s), ~\mathcal{F}^1_1(s)$ \cite{joao_semi}. The lower bound should converge at some point with truncations of the sum over the basis and spins $\ell$--see \cite{boot_white} for the primal bootstrap algorithm. The convergence is visible in our numerics.

The FESRs have errors due to the QCD parameters, mainly from the dimension-four gluon condensate and the dimension-six quark condensates. For convenience, we introduce the following notation for the FESRs
$$
F_n\equiv\frac{1}{s_0^{1+n}}\int_{4}^{s_0}t^n  \frac{\mathrm{Im}\Pi(t)}{(2\pi)^4}dt, ~n=0,1,2\,, ~s_0=1.19 \mathrm{~GeV}^2\,.
$$
Since the FESRs have errors, we naively can't put them as equality. Instead, we must put them as inequality up to a tolerance. A weak possible choice of tolerance is the error, namely 
$
 (\text{mean}-\text{error})<F_n<(\text{mean}+\text{error})\,
$.

 Unitary condition \eqref{semimat} implies that all the principle minors of the matrix $B(s)$ are non-negative, resulting in a simple condition $\rho^1_1(s)\geq |\mathcal{F}^{1}_1(s)|^2$ upon considering the bottom-right minor. Solely using this condition and the FESRs, it is possible to achieve a lower bound of  $630.7^{+3}_{-3}\times 10^{-10}$, which is already better and comparable with the  conservative bound $623\times 10^{-10}$ in \cite{qcdpaper} obtained using FESRs and positivity (considering the smallest possible lower bound due to errors). The full condition \eqref{semimat} and the partial wave unitarity $|S_\ell^I(s)|\leq 1$ together improves the lower bound to $680.0^{+3}_{-3}\times 10^{-10}$. The theory of pion well approximates the low energy QCD due to the chiral symmetry breaking. We use the tree level $\chi$PT to capture the low energy physics. These barely improve the bound (adds half to the third significant digit), but we impose these for completeness. The lower bound is now $680.5^{+3}_{-3}\times 10^{-10}$. Now adding with the charmonium and bottomonium resonance contributions \cite{charm1}, we reach our conservative bound Min$[a_\mu^\text{LO-HVP}]=
688.4^{+3}_{-3}\times 10^{-10}
$.  Combining with other precisely calculated standard model (SM) contributions \cite{th1}, we find the conservative lower bound
$$
a_\mu^\text{bootstrap-min-conservative}=11659176.3^{+3}_{-3}\times 10^{-10}\,.
$$
The prediction from the SM \cite{th1}
$ 
a_\mu^{\text{SM}}= 11659181.0^{+4.3}_{-4.3}\times 10^{-10}\,,
$ within the error bars saturates our lower bound.

\begin{figure}[hbt!]
     \centering
         \includegraphics[scale=0.4]{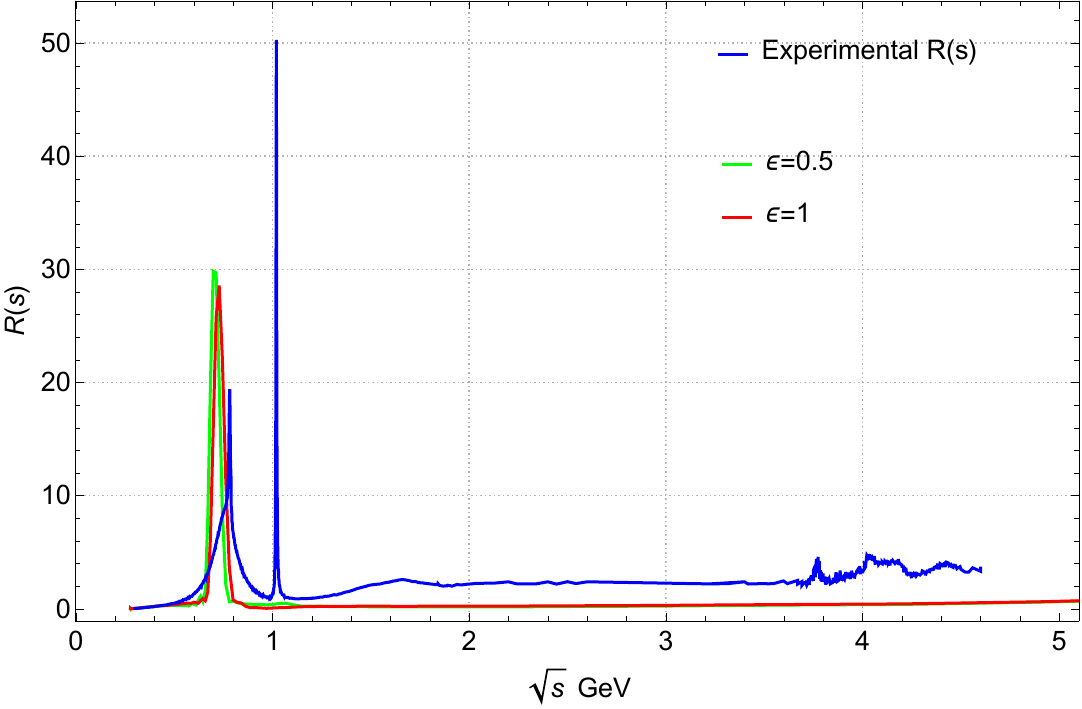}
        \caption{Comparison of bootstrap spectral density with the experimental hadronic cross ratio data.}
        \label{fig:expdata}
\end{figure}

We compare the extremal spectral density with the experimental hadronic cross-ratio data by plotting $12\pi \text{Im}\Pi(s)=R(s)$. For $\rho$--resonances appearing above $\sqrt{s}= 0.7$, bootstrap shows similar features of the location of the peak in experimental hadronic cross-ratio data \cite{PDG2022}-- see figure \eqref{fig:expdata}. The minimisation process returns a spectral density which corresponds to the lower bound. Among all the admissible spectral densities satisfying the bootstrap constraints, namely unitarity, analyticity, crossing symmetry and the FESRs, figure \eqref{fig:expdata} shows the extremal solution that is, the one that saturates the lowest possible value of the hadronic contribution to the muon anomaly. We refer to this as the extremal spectral density, which is generally not expected to match the experimental spectral density. It is not fitted to any data; rather, it is the unique result of minimising and imposing the bootstrap constraints. The sole input from the Standard Model (SM) in our approach is the FESRs, and other constraints from non-perturbative SM can lead to better matching with relevant data used in \cite{th1}. We refer \cite{th1} for the data set used in SM data-driven computation. The spectral function associated with the extremal solution exhibits a localised structure, with most of its support concentrated near a single peak. This differs significantly from the SM data-driven evaluation in \cite{th1}, which incorporates precise input from pion form factors and additional low-energy hadronic contributions, which the bootstrap spectral density doesn't pick up.  

To demonstrate, we plotted the bootstrapped data for truncation ($P=10$) in the computation of $S^I_\ell(s)$ with partial wave unitarity imposed up to $\ell=9$ and the truncation in the basis for $\rho^1_1(s)$ at $95$----refer to next section for truncation in the basis and spins. Data with or without imposing $\chi$PT are almost identical. For these data, the peak position is about $\sqrt{s}=0.73$, corresponding to $\rho$ mass.  

 The lower bound, comparison of the bootstrap spectral density and the $\rho$ mass is impressive as bootstrap results, even with the poor choice of the tolerance. Determining the $\rho$ mass using the QCD sum rules is not novel. Indeed, the literature has long shown how the FESRs can be used to extract the mass of the rho resonance, provided one assumes the existence of such a resonance in the spectral density. Our point, however, is that the appearance of the $\rho$ resonance in our bootstrap-based minimisation of $a_\mu^{\text{had}}$ is notable because we do not impose any resonance structure a priori. This contrasts with traditional sum-rule approaches, which often model the spectral density explicitly as: Im$\Pi(s)\sim\delta(s - m_\rho^2) + \text{continuum}$, or use Breit-Wigner or Gounaris–Sakurai forms, thereby building in the assumption of a rho-like state from the outset. Within our bootstrap setup, we do not input or assume any specific resonance shape or location, yet the $\rho$-like peak emerges dynamically in the solution that minimises $a_\mu^{\text{had}}$. This is why we describe it as an ``impressive bootstrap result". However, a slight improvement in the tolerance for FESRs will lead to an improved lower bound incompatible with SM prediction. The improvement due to the tolerance is evident because the mentioned QCD condensates need to be better determined.  We show the dependence on the tolerance of the lower bound by considering $ (\text{mean}-\e \times \text{error})<F_n<(\text{mean}+\e\times\text{error})$ and vary $\e$ from $1$ to $0$ in table \eqref{tab:e}.  The the average over the tolerance from light quark contributions is
$708.5^{+1.6}_{-1.6}$ and adding with the charmonium and bottomonium resonance contributions, we reach an average lower bound Min$[a_\mu^\text{LO-HVP}]=
716.43^{+1.6}_{-1.6}\times 10^{-10}
$
and adding with the other extensively calculated SM contributions, we find
\be 
a_\mu^\text{bootstrap-min-average}=11659204.3^{+1.6}_{-1.6}\times 10^{-10}\,.
\ee
This is incompatible with the SM prediction while saturated by the lattice evaluation \cite{lat2} $a_\mu^\text{lattice}=11659201.9(3.8)\times 10^{-10}$ and  the measured value $
a_\mu^{\text{exp}}= 11659208.9^{+6.3}_{-6.3}\times 10^{-10}\,,
$ within the error bars. Figure \eqref{fig:bootmuong} summarises our findings. 

The upper bound does not show an apparent convergence, so we avoid revealing the details.

\section{Bootstrap implementations}
We show the details of the numerical implementation of the bootstrap. We write a suitable ansatz for the pion partial waves, form factor and spectral density. 
The pion partial waves $S_\ell^I(s)=1+i \pi \sqrt{\frac{s-4}{s}}f_\ell^I(s)$ are given by
\be 
f_\ell^I(s)=\frac{1}{4}\int_{-1}^1dx P_\ell(x)M^{(I)}\left(s,t=\frac{(s-4)(x-1)}{2}\right)\,,
\ee
where the isospin $I$ channel amplitudes are
\be
\begin{split}
&M^{(0)}=3A(s|t,u)+A(t|s,u)+A(u|t,s)\,,\\
&M^{(1)}=A(t|s,u)-A(u|t,s)\,,\\
&M^{(2)}=A(t|s,u)+A(u|t,s)\,.\\
\end{split}
\ee
The crossing symmetry and analyticity of $A(s|t,u)$ implies the following ansatz \cite{qcdboot1},
\be\label{eqAstu}
\begin{split}
A(s|,t,u)=&\sum _{n=1}^{P} \sum _{m=1}^n a_{n m}\left(\eta_t^m\eta_u^n+\eta_t^n \eta_u^m\right)\\
&+\sum _{n=0}^{P} \sum _{m=0}^{P} b_{n m}\left(\eta_t^m+\eta_u^m\right) \eta_s^n\,,
\end{split}
\ee
where $\eta_z=\frac{\left(\sqrt{4-4/3}-\sqrt{4-z}\right)}{\left(\sqrt{4-4/3+\sqrt{4-z}}\right)}$ and we truncate the sum upto $P$. 
The analyticity of the spectral density and form factor implies the following ansatz \cite{joao_semi},
\be\label{ansatzrhoF}
\begin{split}
&\rho_1^1(s)=-\sum_{n=1}^{N} d_n \sin\left(n \arccos\left(\frac{8}{s}-1\right)\right),\\
&F(s)=\sum_{n=0}^{N} b_n \left(\frac{\sqrt{4}-\sqrt{4-s}}{\sqrt{4+\sqrt{4-s}}}\right)^n \,.
\end{split}
\ee
Note that $b_0=1$ because of  $F(s=0)=1$. 

After writing down the ansatz, we impose the FESRs $F_n$ for $n=0,1,2$. The choice of the $s_0$ is crucial. The lower the value of $s_0$, the better, the lower bound, as was pointed out in \cite{qcdpaper}. However, we can't go arbitrarily low in $s_0$. Below $s_0=1.19$ GeV${}^2$, the strange quark FESRs start to violate simple positivity inequality derived from Holder's inequality \cite{qcdpaper}. Hence, we stop at $s_0=1.19$ GeV${}^2$ --see appendix \ref{ap:qcdsumrules} for details. In our approach, $s_0 = 1.19\,\text{GeV}^2$ should not be interpreted as a cut-off on the QCD contributions. As emphasised in \cite{qcdpaper}, and confirmed by our analysis, lowering $s_0$ improves the strength of the lower bound. We adopt the value $s_0 = 1.19\,\text{GeV}^2$ following \cite{qcdpaper}, even though they go as low as $s_0 = 1.09\,\text{GeV}^2$ for the up and down quark contributions. While \cite{qcdpaper} used only the positivity of the spectral density to constrain the hadronic contribution, our work builds upon and significantly improves it by imposing a more general version of unitarity—not just positivity—as well as crossing symmetry and analyticity. These are the full set of bootstrap constraints, as detailed in the bootstrap strategy. Our findings indicate that the bootstrap constraints are strictly stronger than positivity alone. The resulting lower bounds are thus as rigorous as those derived from the FESRs, but enhanced by the additional consistency conditions imposed by the bootstrap approach. The FESRs, along with the errors coming from the QCD parameters (at $s_0=1.19$ GeV${}^2$)

\be
\begin{split}
F_0&=0.0000416772_{-0.00000000105807}^{+0.00000000260880}\,, \\
F_1&=0.0000186454\pm 6.4034\times 10^{-8}\,,\\
F_2&=9.17113\times 10^{-6}\pm 5.6487\times 10^{-7}\,.
\end{split}
\ee
\\
The sum rules are given in the appendix \ref{ap:qcdsumrules}. The dimension-four gluon condensate $\langle \a G^2\rangle$, vacuum saturation constant ($\kappa$) provide dominant contributions to the errors \cite{qcdpaper}, where $\kappa$ expresses dimension-six quark condensates as products of dimension-three quark condensates,  $\alpha_s \langle (\bar{n}n)^2 \rangle=\kappa \alpha_s \langle \bar{n}n \rangle^2$ \cite{refpara_k}. For the convenience of determining tolerance, we considered an error for the first sum rule coming from strange quark mass, even though it has no visible effect in numerics.

Since the FESRs have errors, we naively can't put them as equality. Rather, we must put them as inequality up to a tolerance. A weak possible choice for the tolerance is the error, namely $ (\text{mean}-\text{error})<F_n<(\text{mean}+\text{error})$. 

\subsection{Step by step bootstrap }
We illustrate the bootstrap implementation in three steps. Firstly, we consider the simplest bootstrap constraint between form factor and spectral density. The result from the first step is comparable to known literature. In the second step, we consider the full bootstrap conditions and increasing partial wave unitary constraints spin by spin. Thirdly, we consider chiral symmetry breaking, which improves the numerics slightly.

\begin{figure}[hbt!]
     \centering   
         \includegraphics[scale=0.25]{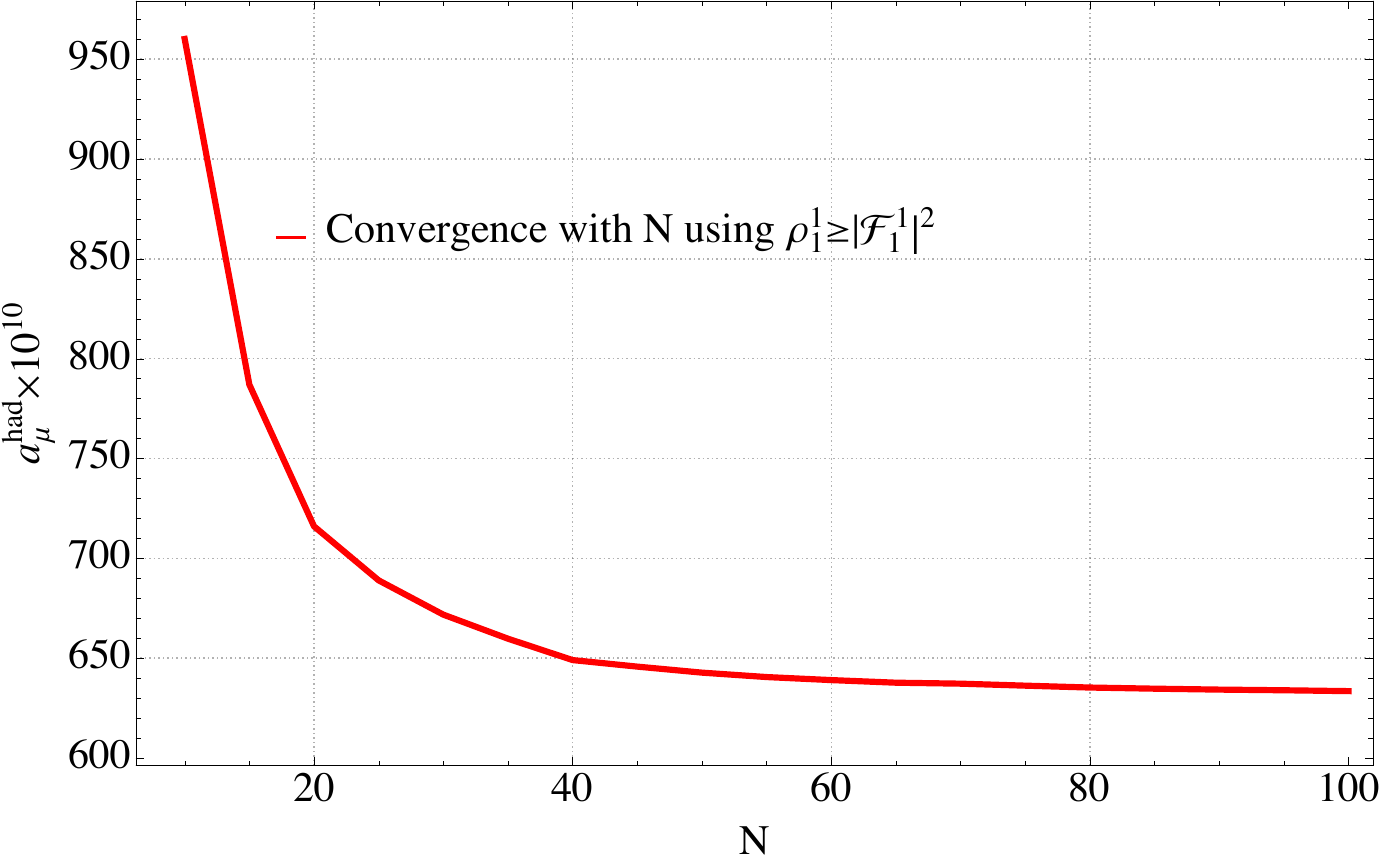}
        \caption{Convergence of numerics with the number of basis elements $N$ for the simplest condition $\rho^1_1(s)\geq |\mathcal{F}^{1}_1(s)|^2$.}
        \label{fig:rhoF}
\end{figure}

\textit{\textcolor{Blue}{Step 1$\setminus$} Simplest condition for form factor and spectral density:}
Unitary condition \eqref{semimat} implies that all the principle minors of the matrix $B(s)$ are non-negative, resulting in a simple condition $\rho^1_1(s)\geq |\mathcal{F}^{1}_1(s)|^2$ upon considering the bottom-right minor. Solely using this condition and $ (\text{mean}-\text{error})<F_n<(\text{mean}+\text{error})$, it is possible to achieve a minimum for  $a_\mu^\text{LO-HVP}$ as demonstrated in figure \eqref{fig:rhoF}. 
The extrapolation for large number of basis elements ($N$) gives Min$[a_\mu^\text{LO-HVP}]=630.7^{+3}_{-3}\times 10^{-10}$. We do extrapolations for large $N$ with different models and average the errors and mean values \cite{andreafit}. Since convergence at the third significant digit is evident, we discarded models that significantly deviated from these values. 
A reasonable comparison for the number Min$[a_\mu^\text{LO-HVP}]=630.7^{+3}_{-3}\times 10^{-10}$ can be found in \cite{qcdpaper}. In \cite{qcdpaper}, a two-sided bound on $a_\mu^\text{LO-HVP}$ was derived using positivity of the spectral density and  FESRs for each quark section utilizing Holder's inequalities. For lower bound, the authors noticed that simple form of Kernel $K(t)$ enables to write  $a_\mu^\text{LO-HVP}\geq 0.83\times  \frac{4 \a^2 m_\mu^2}{3\pi} \times \int_{4m_\pi^2}^{\infty}\frac{\text{Im}\Pi(t)}{t^2}$ and FESRs puts a lower bound on $\int_{4m_\pi^2}^{\infty}\frac{\text{Im}\Pi(t)}{t^2}$. Considering errors for FESRs coming from gluon condensate $\langle \a G^2\rangle$, vacuum saturation constant ($\kappa$) they arrive at conclusion that $a_\mu^\text{LO-HVP}> 657_{-34}^{+34}\times 10^{-10}$. Since we are using the FESRs as inequalities due to errors and minimization process picks up the lowest of the bound, hence correct number we should compare is $a_\mu^\text{LO-HVP}> 623\times 10^{-10}$, which is in good agreement with our lower bound Min$[a_\mu^\text{LO-HVP}]=630.7^{+3}_{-3}\times 10^{-10}$ achieved using the simplest condition $\rho^1_1(s)\geq |\mathcal{F}^{1}_1(s)|^2$ and the FESRs.

\begin{figure}[hbt!]
     \centering
         \includegraphics[scale=0.25]{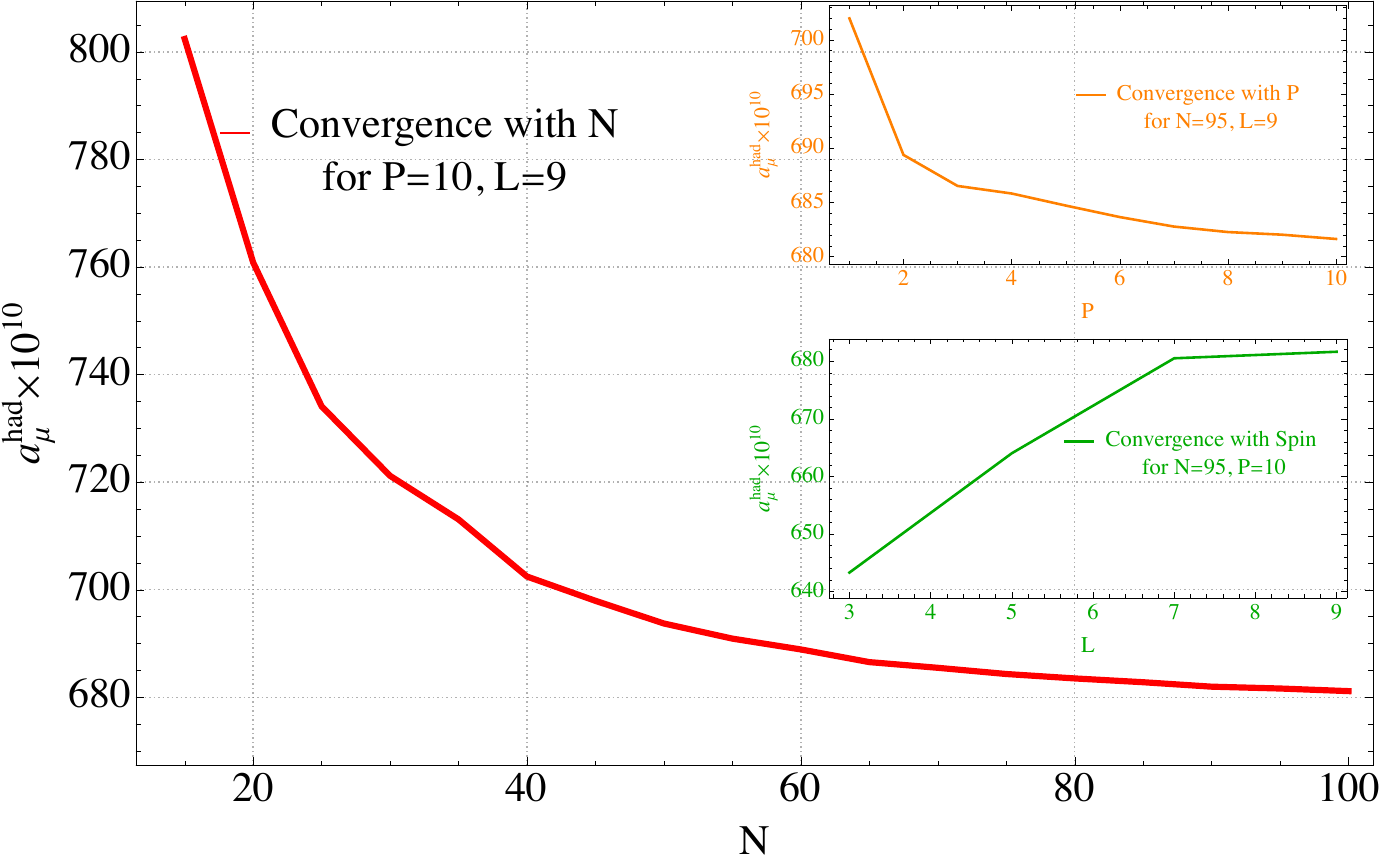}
        \caption{Convergence of numerics with $N, L, P$. We impose partial wave unitarity up to $\ell=1,3,5,\dots L$ for isospin $I=1$ and $\ell=0,2,4,\dots L-1$ for isospin $I=0,2$, $P$ is the truncation of the sum for ansatz for $A(s|t,u)$ in \eqref{eqAstu} and $N$ is the truncation in \eqref{ansatzrhoF}.}
        \label{fig:LPconvergence}
\end{figure}

\textit{\textcolor{Blue}{Step 2$\setminus$} Comprehensive constraints for form factor, spectral density and partial waves:}
We now focus on full numerics after demonstrating a simple form of numerics and a successful comparison. We implement the condition \eqref{semimat} by converting the $B(s)$ matrix into a $6\times 6$ matrix \cite{joao_semi} with an equivalent condition $\left(
\begin{array}{cc}
\text{Re}B(s) & -\text{Im}B(s) \\
\text{Im}B(s)  & \text{Re}B(s) 
\end{array}
\right) \succeq 0$ using SDPB solver \cite{sdpb}.  We impose the partial wave unitarity $|S_\ell^I(s)|\leq 1$ upto spin $L$, namely $\ell=1,3,5,\dots L$ for isospin $I=1$ and $\ell=0,2,4,\dots L-1$ for $I=0,2$. We remind the reader that the truncation in the number of basis elements is denoted by $N$ in eq \eqref{ansatzrhoF} and $P$ in eq \eqref{eqAstu}. The minimum should stabilize at some point with $N,L,P$--see \cite{boot_white} for the primal bootstrap algorithm. The convergence with $N, L, P$ are shown in figure \eqref{fig:LPconvergence}. Truncating the spin at $L=9$ and $P=10$ does not alter the third significant digits. Hence, throughout our analysis, we use these truncations. The convergence with $N$ is evident in figure \eqref{fig:LPconvergence}. We do extrapolations for large $N$ with different strategies (for example $a+\frac{b}{N^2}$ with $a=682.1\pm 2.8$, $a+b \exp(-0.07 N)$ with $a=682.9\pm 2.4$ e.t.c) and average the errors and mean values following \cite{andreafit}. For light quark contribution, the final bound in the second step is $680.0^{+3}_{-3}\times 10^{-10}$, which shows the improvement from full unitarity. 

\textit{\textcolor{Blue}{Step 3$\setminus$}  Imposing chiral symmetry breaking:}
The theory of pion well approximates the low energy QCD due to chiral symmetry breaking. We use tree-level $\chi$PT to capture the low energy physics. These barely improve the bound (adds half to the third significant digit), but we impose these for completeness. The tree-level partial waves are 
\be
\begin{split}
&f^0_{0,\text{tree}}(s)=\frac{2	}{\pi}\frac{2 s-1}{ 32 \pi  f_{\pi}^2},~ f^1_{1,\text{tree}}(s)=\frac{2	}{\pi}\frac{s-4}{96 \pi f_{\pi}^2} , \\
&f^2_{0,\text{tree}}(s)=\frac{2	}{\pi}\frac{2-s}{32 \pi  f_{\pi}^2}\,.
\end{split}
\ee
For $0\leq s \leq 4$, we impose $|f^I_{\ell}(s)-f^I_{\ell,\text{tree}}(s)|<3\times 10^{-2}$. The tolerance $3\times 10^{-2}$ is dictated by the 2-loop answer which same as in \cite{Heboth}. For example $f^0_{0,\text{tree}}$ differs maximum at $s=4$ with 2-loop answer which is about $25\%$, hence we use tolerance of $30\%$. We impose these inequalities for $0\leq s\leq 4$ with a spacing $1/2$. We observed that reducing the spacing to $1$ does not change the answers to the 4th significant digits.

\begin{figure}
     \centering
         \includegraphics[scale=0.25]{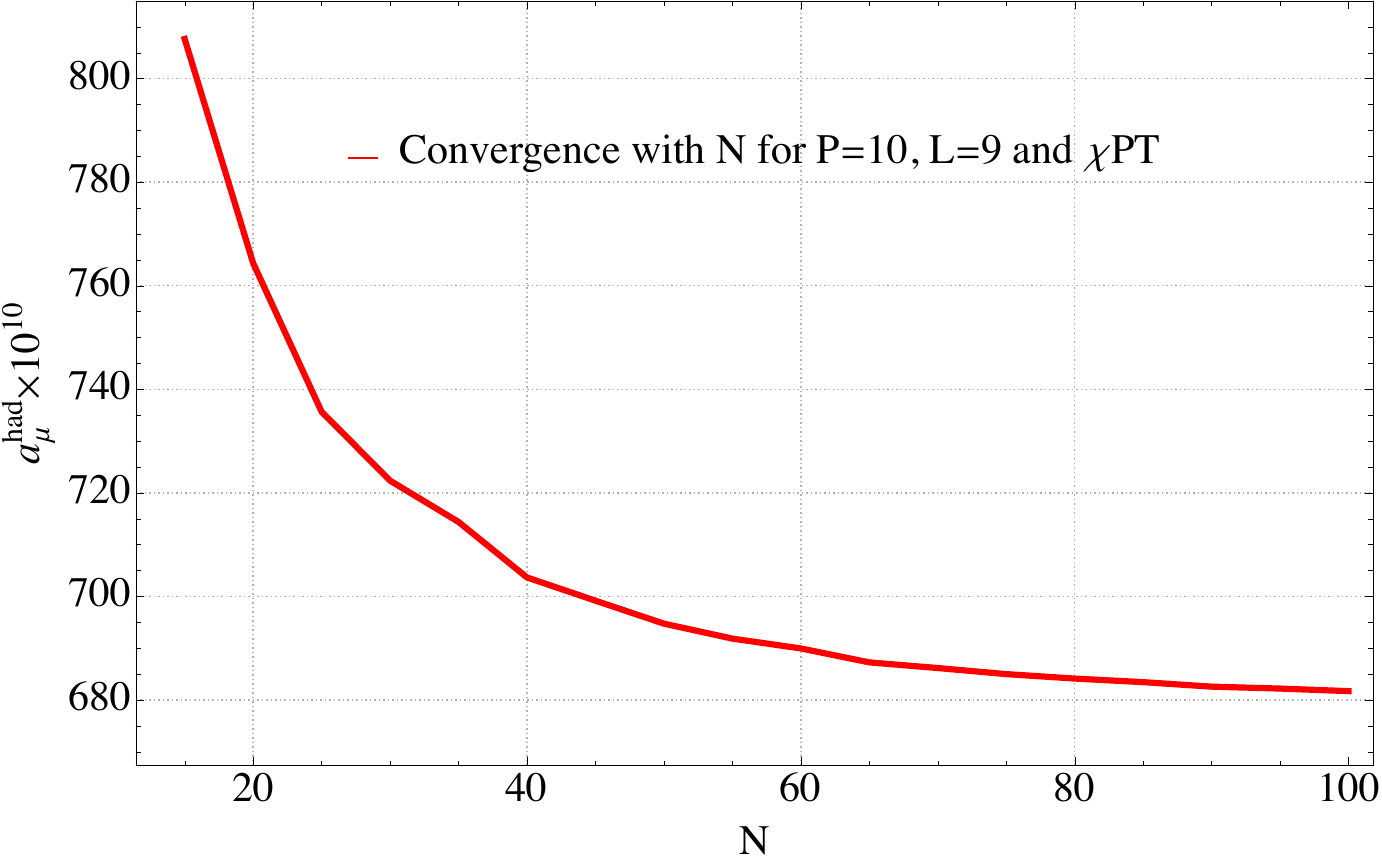}
        \caption{Convergence of numerics with $N$,  the truncation in \eqref{ansatzrhoF} imposing $\chi$PT.}
        \label{fig:N_chipt}
\end{figure}

The convergence for the lower bound is shown in figure \eqref{fig:N_chipt}. The extrapolated value for large N is $680.5^{+3}_{-3}$. Now adding with the charmonium and bottomonium resonance contributions \cite{charm1}, we reach our final bound Min$[a_\mu^\text{LO-HVP}]=
688.4^{+3}_{-3}\times 10^{-10}
$

\subsection{Dependence on tolerance}
\begin{figure}
     \centering
         \includegraphics[scale=0.25]{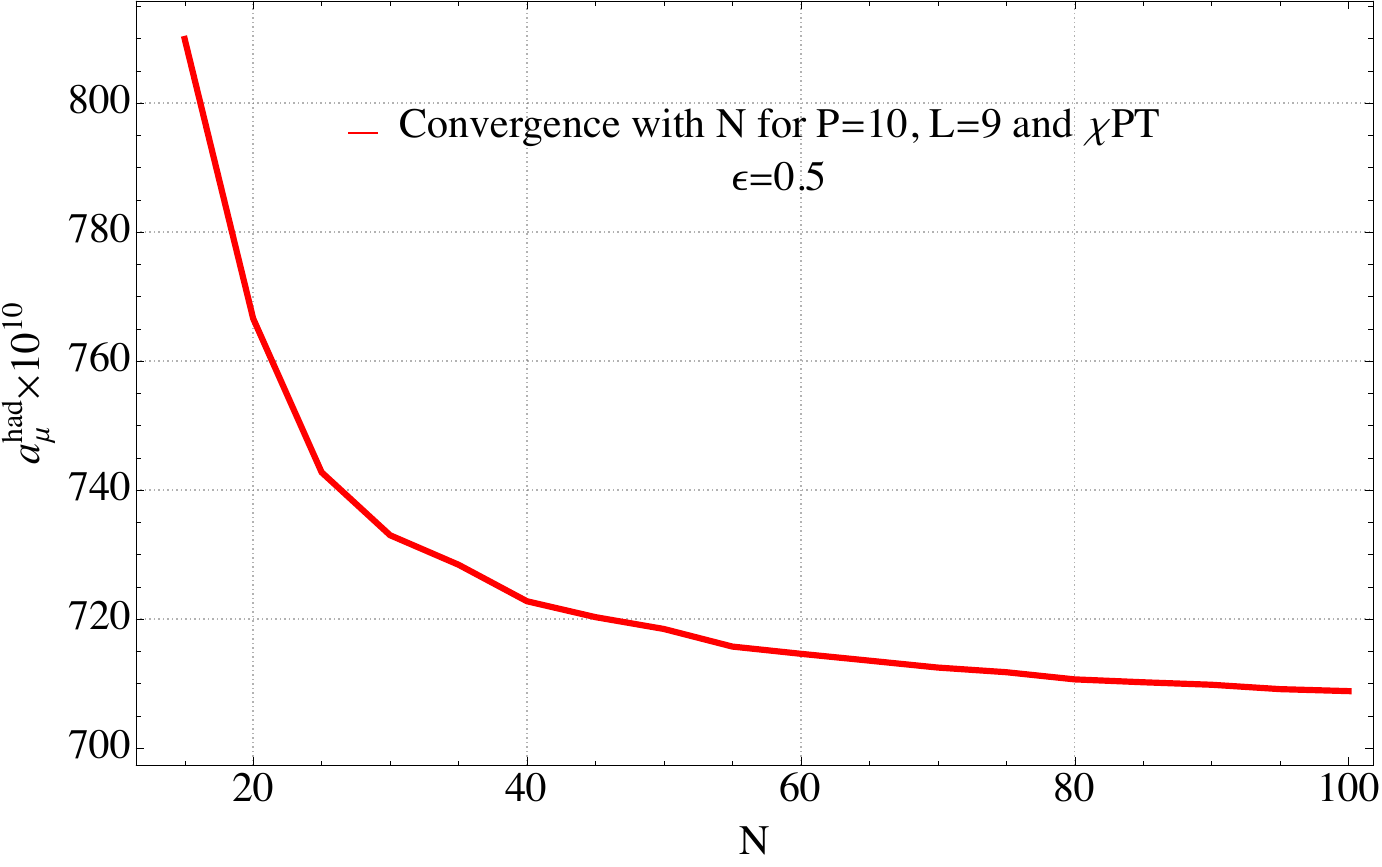}
        \caption{Convergence of the numerics with $N$, the truncation in \eqref{ansatzrhoF} considering the $\chi$PT for $\e=1/2$.}
        \label{fig:N_sigma}
\end{figure}

In this section, we show the dependence on the tolerance of the FESR inequalities, namely we consider $ (\text{mean}-\e \times \text{error})<F_n<(\text{mean}+\e\times\text{error})$ and vary $\e$ from $1$ to $0$. Since the results have already converged around $N=95$, we show the dependence of Min$a_\mu^\text{had}$ on $\e$ in table \eqref{tab:e}. We find that the average over these 11 choices of $\e$ is $710$, which coincides with the value at $\epsilon = 1/2$ (differing only by one). We show the convergence in figure \eqref{fig:N_sigma} for completeness. The extrapolation for a large number of basis elements $N$ gives  $708.5^{+1.6}_{-1.6}$ and combining the charmonium and bottomonium resonance contributions, we reach our average bound Min$[a_\mu^\text{LO-HVP}]=
716.43^{+1.6}_{-1.6}\times 10^{-10}
$.
\begin{table}
\begin{center}
\begin{tabular}{CCCCCCCCCCC}
\hline
~~~~~~~~~~~\epsilon  = 1& 0.9 & 0.8 & 0.7 & 0.6 \\

  {\tiny a_{\mu }^{\text{had}}\times 10^{10}} = 681.6 & 687 & 692 & 697 & 703  \\
\hline
~~~~~~~~~~~~~\epsilon=0.5 & 0.4 & 0.3 & 0.2 & 0.1 & 0 \\

 {\tiny {\tiny a_{\mu }^{\text{had}}\times 10^{10}}}= 709 &  716 & 722 & 729 & 735 & 742 \\
 \hline\
\end{tabular}
\end{center}
\caption{{Dependence on tolerance$=\e\times$error. The average over tolerance is at $\e=1/2$.} }
\label{tab:e}
\end{table}


The bootstrap solution for the $\e=1/2$ lower bound corresponds to $\langle \alpha G^2 \rangle=0.06315 \, \mathrm{GeV}^4$
, $\kappa$ $=3.47$, while the known literature values are $0.0649 \pm 0.0035 \, \mathrm{GeV}^4$
, $3.22 \pm 0.5$, respectively--see tables in appendix \ref{ap:qcdsumrules}. Note that corresponding lower bound is saturated by lattice data and is within the errorbars of exp. The agreement of the average lower bound with the lattice and the  measured value suggests that these could be the potential numbers for $\langle \alpha G^2 \rangle, ~\kappa$. For future determination these numbers can serve as a benchmark points.  

\subsection{{Constraints on condensates}}
\begin{figure}
     \centering
         \includegraphics[scale=0.37]{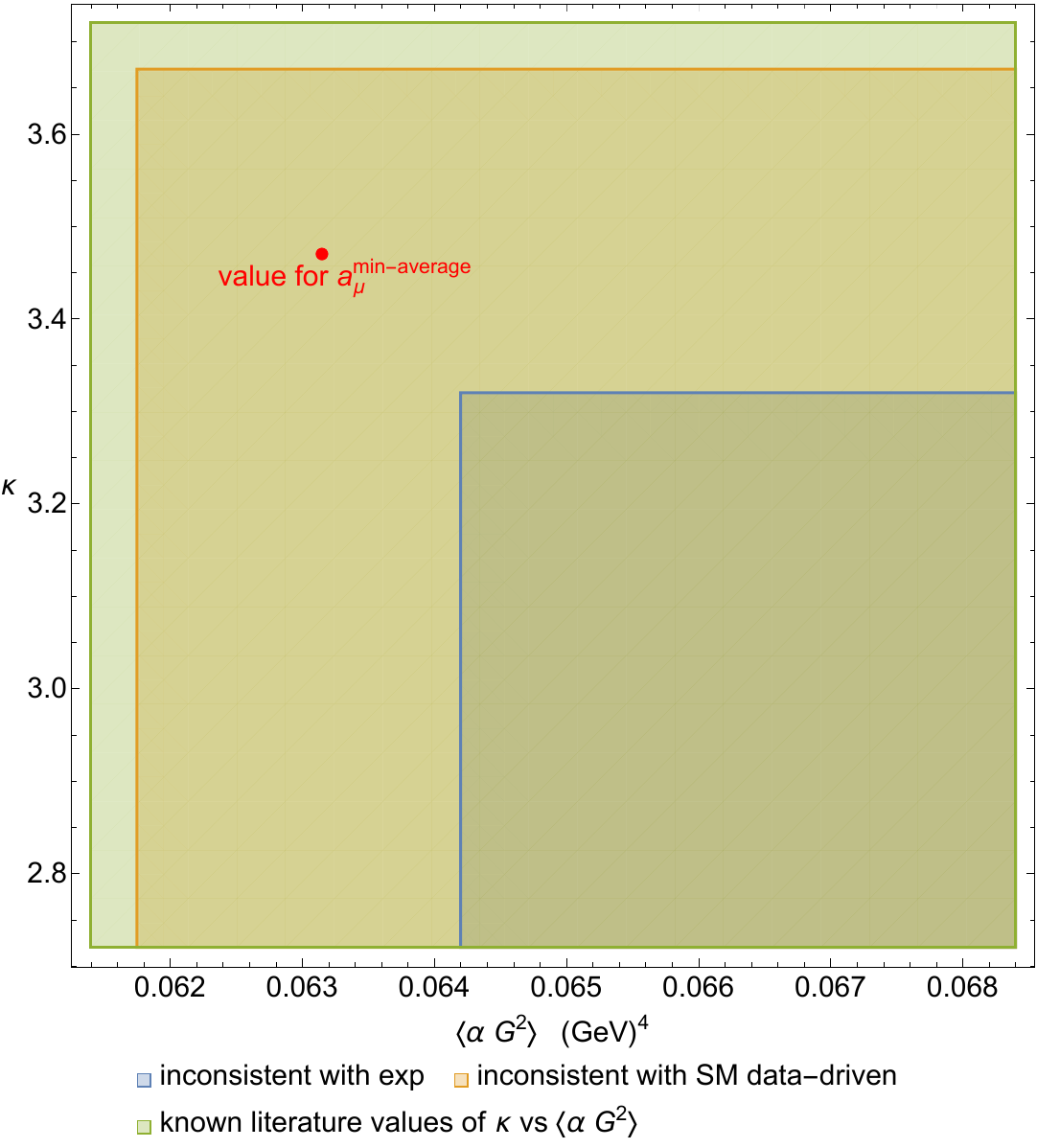}
        \caption{The green region corresponds to the known literature values of \(\kappa\) and \(\langle \alpha G^2 \rangle\). Yellow region is inconsistent with the SM data-driven prediction of the muon anomaly while blue region is ruled out by  the measured value. Note that the intersection of the excluded SM and experimental regions forms a characteristic \(\big{\ulcorner}\)-shaped area. The average of our lower bounds on muon anomaly lies approximately at the center of this corner. We propose this location as a benchmark for future computations/searches for condensates. We thank the anonymous referees for this suggestion.}
        \label{fig:condensates}
\end{figure}

From figure \eqref{fig:bootmuong}, it is evident that some choice of the tolerances are ruled out by the measured values of the muon anomaly, roughly the tolerance smaller than $\epsilon=0.2$ are ruled out by measured value of the muon anomaly, while smaller than $\epsilon=0.9$ are inconsistent with the SM data-driven prediction. Roughly speaking, the experimental constraint requires that
$
\text{mean} - 0.2 \times \text{error} < F_n < \text{mean} + 0.2 \times \text{error}
$ are ruled out. However, the situation is actually more subtle and informative than this simple inequality suggests. Since we are working with three FESRs, and  the dominant uncertainties arise from the \(n=1\) and \(n=2\) moments, which involve the condensates \(\kappa, \langle \alpha G^2 \rangle\). We can examine more precisely which regions of the condensate parameter space are ruled out.
The \(n=1\) FESR saturates the lower limit \(\text{mean} - 0.2 \times \text{error}\),  
while the \(n=2\) FESR saturates the upper limit \(\text{mean} + 0.2 \times \text{error}\).
This implies that following regions 
$
F_1 > \text{mean} - 0.2 \times \text{error}
\quad \text{and} \quad
F_2 < \text{mean} + 0.2 \times \text{error}
$
are excluded.
To explain this more clearly:  
Suppose
$
F_1 = \text{mean} - 0.2 \times \text{error}
\quad \text{and} \quad
F_2 < \text{mean} + 0.2 \times \text{error}.
$
In this case, the lower bound on the muon anomaly would be stronger compared to the situation where both integrals \textit{exactly} saturate their respective bounds. Similarly, if
$
F_1> \text{mean} - 0.2 \times \text{error}
\quad \text{and} \quad
F_2\,  < \text{mean} + 0.2 \times \text{error},
$
the lower bound would become even stronger.
This implies that the region is ruled out
$
(0.0649 - 0.2 \times 0.0035)~\mathrm{GeV}^4 < \langle \alpha G^2 \rangle
$
if
$
\kappa < 3.22 + 0.2 \times 0.5,
$
and vice versa.  
It is important to note that these bounds are not independent of one another.
At present, we are unable to determine an upper bound on \(\langle \alpha G^2 \rangle\) or a lower bound on \(\kappa\) — nor independent two-sided bounds  because only one experimental constraint from \(a_\mu^{\text{exp}}\) is available. It would be interesting to explore whether additional theoretical or experimental constraints could establish two-sided (and independent) bounds on these condensates.
This situation is illustrated in figure~\eqref{fig:condensates}. Once can be more precise and use three different tolerance for three different FESRs, but quantitative picture will be same. 

\subsection*{Conclusion}
We conclude that unitarity, analyticity, crossing symmetry and the FESRs can establish a lower bound on $a_\mu^\text{LO-HVP}$, adding with rest of the extensively calculated SM contribution, we reach a lower bound $a_\mu^\text{min}$.  Our bootstrap results are consistent with the measured values of the muon anomaly. The bootstrapped spectral density shows a features like $\rho$-resonance peak similar to experimental hadronic cross-ratio data, proving a bootstrap prediction for $\rho$-meson mass, further underscoring the robustness of our approach.

\section*{Acknowledgments} 
We thank Bobby Samir Acharya, Subham D. Chowdhury, Paolo Creminelli, Atish Dabholkar, Ehsan Ebrahimian, Aditya Hebbar, Joan Elias Miro, Andrea L. Guerrieri, Mehmet A. Gumus, Thomas G. Steele, Aninda Sinha, Shaswat Tiwari, and Alexander V. Zhiboedov for their helpful discussions. We would also like to thank SISSA, Trieste for supporting with their powerful clusters. I have received support from the European Research Council, grant agreement n. 101039756.

\appendix


\section{Further details on the numerics}
\subsection{Muon anomaly and the positive semi-definite matrix}
The leading hadronic muon anomaly is given by
\begin{equation}
    a_\mu^\text{LO-HVP} = \frac{4\alpha^2}{\pi} \int_{4m_\pi^2}^{\infty}\frac{K(t) \text{Im}\Pi(t)}{t} \, dt\,,
\end{equation}
We want to determine the minimum of this integral imposing the unitary condition among Im$\Pi(t)$, pion partial wave $S_1^1$ and form factor given  by\cite{joao_semi,Heboth}

\be\label{semimata}
B(s)\equiv\left(
\begin{array}{ccc}
1 & S_1^1(s) & \mathcal{F}_1^1(s) \\
S_1^{1*}(s) & 1 & \mathcal{F}_1^{1*}(s) \\
\mathcal{F}_1^{1*}(s) & \mathcal{F}_1^{1}(s) & \rho_1^1(s)
\end{array}
\right) \succeq 0, \quad s > 4\,,
\ee
where $\rho^1_1(s)\times \frac{(2\pi)^4}{s}=\text{Im}\Pi(s)$ and $\mathcal{F}_1^1(s)=\frac{\sqrt{\frac{4 \pi }{3}} \left(\frac{s-4}{4}\right)^{3/4}}{\left(8 \pi ^3\right) \sqrt[4]{s}}F(s) $, 
with $F(s)$ being some vector form factor normalized as $F(0)=1$. This positive semi-definite matrix implies all the minors should be positive including determinant. We get three constraints \cite{joao_semi}
\begin{enumerate}
\item From the right-bottom minor:\be \rho_1^1(s) \geq  |\mathcal{F}_1^1(s)|^2.\ee
\item From the top-left minor: \be|S_1^1(s)| \leq  1.\ee
\item From the determinant of $B(s)$:
\be\rho _1^1 \left(1-\left| S_1^1\right|^2\right)-2 \left| \mathcal{F}_1^1\right|^2+S_1^1 \left(\mathcal{F}_1^1{}^*\right){}^2+S_1^1{}^*(\mathcal{F}_1^{ 1})^2\geq 0.\ee
\end{enumerate}

Note that the first condition is even stronger than sole positivity $\rho^1_1>0$ or Im$\Pi>0$. Second condition is the usual partial wave unitarity of the pion partial waves. Third constraint very non-trivially relates Im$\Pi,~ \mathcal{F}^1_1$ and $S_1^1$ with each other. {These constraints are stronger and non-trivial than positivity Im$\Pi>0$. Note that in \cite{qcdpaper} Im$\Pi>0$ positivity is used. Therefore upon using the non-trivial unitarity conditions stated above, we  expect stronger bound than \cite{qcdpaper}.}

In computer, we implement the condition \eqref{semimata} by converting the $B(s)$ matrix into a $6\times 6$ matrix as describe in \cite{joao_semi} with an equivalent condition $\left(
\begin{array}{cc}
\text{Re}B(s) & -\text{Im}B(s) \\
\text{Im}B(s)  & \text{Re}B(s) 
\end{array}
\right) \succeq 0$. These are complicated numerics needs better precision. We solve them using SDPB solver \cite{sdpb} a extensively used tools in S-matrix bootstrap to solve this kind of problem. 
\subsection{Pion partial waves: crossing symmetry, analyticity and unitarity}
Pions have partial waves $S_\ell^I$ for spins $\ell=0,1,2,3\dots$ and iso-spins $I=0,1,2$. Note that appearance of $S_1^1$ is very non-trivial, it puts  non-trivial constrains on $\rho^1_1$. The $S_1^1$ satisfies unitarity, and is related to other partial wave coefficients due to crossing and analyticity of pion amplitudes. The pion partial waves $S_\ell^I(s)=1+i \pi \sqrt{\frac{s-4}{s}}f_\ell^I(s)$ are given by
\be 
f_\ell^I(s)=\frac{1}{4}\int_{-1}^1dx P_\ell(x)M^{(I)}\left(s,t=\frac{(s-4)(x-1)}{2}\right)\,,
\ee
where the isospin $I$ channel amplitudes are
\be
\begin{split}
&M^{(0)}=3A(s|t,u)+A(t|s,u)+A(u|t,s)\,,\\
&M^{(1)}=A(t|s,u)-A(u|t,s)\,,\\
&M^{(2)}=A(t|s,u)+A(u|t,s)\,.\\
\end{split}
\ee
The crossing symmetry and analyticity of $A(s|t,u)$ implies the following ansatz \cite{qcdboot1},
\be\label{eqAstu}
\begin{split}
A(s|,t,u)=&\sum _{n=1}^{P} \sum _{m=1}^n a_{n m}\left(\eta_t^m\eta_u^n+\eta_t^n \eta_u^m\right)\\
&+\sum _{n=0}^{P} \sum _{m=0}^{P} b_{n m}\left(\eta_t^m+\eta_u^m\right) \eta_s^n\,,
\end{split}
\ee
where $\eta_z=\frac{\left(\sqrt{4-4/3}-\sqrt{4-z}\right)}{\left(\sqrt{4-4/3+\sqrt{4-z}}\right)}$. The $a_{nm}$ and $b_{nm}$ are coefficients to be optimised and  $P$ is the truncation level of the ansatz.

Using the above ansatz, we can compute the the $S_\ell^I$, which depends on $a_{nm}, b_{nm}$ and truncation $P$. Because of the same parameter dependent of all $S_\ell^I[s,a_{nm}, b_{nm},P]$, they are related/constrainted, which is nothing but consequences of crossing symmetry and analyticity--see \cite{qcdboot1} for details. Hence \be \begin{split}
&|S_\ell^I[s,a_{nm}, b_{nm},P]|\leq 1,\\
 &\text{ for spins $\ell=0,1,2,3\dots$} \text{and iso-spins $I=0,1,2$.} 
\end{split}
\ee 
puts some further constrains on $S_1^1[s,a_{nm}, b_{nm},P]$. 
\subsection{Implementing the positive semi-definite condition}
Usual practice to search for the functions Im$\Pi,~ \mathcal{F}^1_1$ and $S_1^1$ that minimises $a_\mu^\text{had}$ and satisfy \eqref{semimata} is by writing ansatz for Im$\Pi,~ \mathcal{F}^1_1$ and $S_1^1$. Following  \cite{joao_semi} we use the ansatz
\be\label{ansatzrhoF}
\begin{split}
&\rho_1^1(s)=-\sum_{n=1}^{N} d_n \sin\left(n \arccos\left(\frac{8}{s}-1\right)\right),\\
&F(s)=\sum_{n=0}^{N} b_n \left(\frac{\sqrt{4}-\sqrt{4-s}}{\sqrt{4+\sqrt{4-s}}}\right)^n \,.
\end{split}
\ee
where $d_n$ and $b_n$ are coefficients to be optimised and $N$ determines the truncation level of the ansatz. Putting all these together, we get 
\begin{widetext}
\be\label{semimatb}
\begin{split}
B(s, d_n, b_n, a_{nm}, b_{nm})\equiv\left(
\begin{array}{ccc}
1 & S_1^1(s, a_{nm}, b_{nm},P) & \mathcal{F}_1^1(s, b_n, N) \\
S_1^{1*}(s, a_{nm}, b_{nm},P) & 1 & \mathcal{F}_1^{1*}(s, b_n, N) \\
\mathcal{F}_1^{1*}(s, b_n, N) & \mathcal{F}_1^{1}(s, b_n, N) & \rho_1^1(s,d_n,N)
\end{array}
\right)\succeq 0.
\end{split}
\ee
\end{widetext}
and 
\be 
|S_\ell^I[s,a_{nm}, b_{nm},P]|\leq 1.
\ee
These should satisfy for all $s>4$. We discretize $s$ into 200 points adopting from \cite{Sboot3}:
\begin{equation}\nonumber
\begin{split}
&s[j] = \frac{\frac{4}{3} \left(1-\exp \left(\frac{1}{203} i (\pi  j)\right)\right)^2+16 \exp \left(\frac{1}{203} i (\pi  j)\right)}{\left(1+\exp \left(\frac{1}{203} i (\pi  j)\right)\right)^2},\,
\end{split}
\end{equation}
where $j = 1, \dots, 200$. Note that we explicitly showed the dependence on the variables and coefficients. Note that $S_\ell^I(s)$ depends on $s, a_{nm}, b_{nm}$, while  $\mathcal{F}_1^1(s)$ depends on $s,b_n$ and $\rho_1^1(s)$ depends on $s,d_n$, hence the notation $B(s, d_n, b_n, a_{nm}, b_{nm})$.

\subsection{The minimization problem}
We want to solve the optimization problem:

\begin{widetext}
\begin{large}
\begin{equation}
\min_{d_n, b_n, a_{nm}, b_{nm}} a_\mu^\text{LO-HVP}[d_n,N]\,,
\end{equation}
\end{large}
\centering subject to the constraints:
\begin{enumerate}
\item Spectral density unitarity~~~ \begin{large}
$B(s[j], d_n, b_n, a_{nm}, b_{nm}) \succeq 0 \quad \text{for all } j = 1, \dots, 200.
$
\end{large}

\item Partial wave unitarity~~~~~ \begin{large}$
|S_\ell^I(s[j], a_{nm}, b_{nm},P)| \leq  1, \text{ for all } \ell=0,1,2,3\dots, I=0,1,2\,,
$\end{large}
\item FESR sum rules~~~~~~~~~~ \begin{large}
$(\text{mean}-\e \times \text{error})<F_n<(\text{mean}+\e\times\text{error}),~~n=0,1,2.
$
\end{large}
\end{enumerate}
\end{widetext}
Note that $$F_k=\frac{1}{s_0^{1+k}}\int_{4}^{s_0}t^{k-1} \rho_1^1(t)=\frac{1}{s_0^{1+k}}\int_{4}^{s_0}t^k  \frac{\mathrm{Im}\Pi(t)}{(2\pi)^4}dt.$$ There are three different sum rules, we choose one tolerance for all of them. Taking three different choices doesn't change the conclusions of figure 1. 

We implement these semidefinite conditions using the \textbf{SDPB solver} (a specialized software for solving semidefinite problems).

\subsection{Checking for convergence with $N,P$}

Since our basis expansions are truncated at $N$ and $P$, we check if our results stabilize as we increase the $N$ and $P$. We verify whether the minimum remains unchanged beyond a certain $N$ and $P$. If the value fluctuates significantly, we increase $N$ and $P$ until it converges. The lower bound should converge at some point with truncations of the sum over basis and spins $\ell$--see \cite{boot_white} for the primal bootstrap algorithm. Convergence with $N,P, \ell$ is visible in the numerics as shown in figures  3, 4, 5, 6 in the main text . 

\subsection{ Three steps}

We show details of three steps:

\textit{\textcolor{Blue}{Step 1$\setminus$} Simplest condition for form factor and spectral density:}

At this stage, we impose the simplest constraint on the spectral density:
\be
\rho_1^1(s) \geq |\mathcal{F}_1^1(s)|^2,
\ee
which is already stronger than positivity. This translates into the following semi-positive definite matrix condition:

\be
\left(
\begin{array}{cc}
 1 & \mathcal{F}_1^{1*}(s, b_n, N) \\
 \mathcal{F}_1^{1}(s, b_n, N) & \rho_1^1(s,d_n,N)
\end{array}
\right)\succeq 0.
\ee
Additionally, we enforce the Finite Energy Sum Rules (FESRs):
\be
\begin{split}
&(\text{mean}-\e \times \text{error})<F_n<(\text{mean}+\e\times\text{error}).
\end{split}
\ee
The dependence of $a_\mu^\text{LO-HVP}[d_n,N]$ on $N$ is shown in Figure 3 for $\epsilon=1$. Convergence is observed around $N=95$. For completeness, an extrapolation to large $N$ yields the minimal value: Min$[a_\mu^\text{LO-HVP}]=630.7^{+3}_{-3}\times 10^{-10}$. 

Even at this preliminary stage, our results exhibit slight improvements over those reported in \cite{qcdpaper}. In \cite{qcdpaper}, a two-sided bound on $a_\mu^\text{LO-HVP}$ was derived using positivity of the spectral density and  FESRs for each quark section utilizing Holder's inequalities. For lower bound, the authors noticed that simple form of Kernel $K(t)$ enables to write  $a_\mu^\text{LO-HVP}\geq 0.83\times  \frac{4 \a^2 m_\mu^2}{3\pi} \times \int_{4m_\pi^2}^{\infty}\frac{\text{Im}\Pi(t)}{t^2}$ and FESRs puts a lower bound on $\int_{4m_\pi^2}^{\infty}\frac{\text{Im}\Pi(t)}{t^2}$. Considering errors for FESRs coming from gluon condensate $\langle \a G^2\rangle$, vacuum saturation constant ($\kappa$) they arrive at conclusion that $a_\mu^\text{LO-HVP}> 657_{-34}^{+34}\times 10^{-10}$. The errors account for the FESR uncertainties. However, since the weakest bound in their case corresponds to $a_\mu^\text{LO-HVP} > 623 \times 10^{-10}$, which is in good agreement with our lower bound Min$[a_\mu^\text{LO-HVP}]=630.7^{+3}_{-3}\times 10^{-10}$ achieved using simplest condition $\rho^1_1(s)\geq |\mathcal{F}^{1}_1(s)|^2$ and FESRs.

\textit{\textcolor{Blue}{Step 2$\setminus$} Comprehensive constraints for form factor, spectral density and partial waves:}

At this step, we use full constraints namely
\begin{enumerate}
\item Spectral density unitarity \begin{equation}
B(s[j], d_n, b_n, a_{nm}, b_{nm}) \succeq 0 \quad \text{for all } j = 1, \dots, 200.
\end{equation}
\item Partial wave unitarity\begin{equation}
|S_\ell^I(s[j], a_{nm}, b_{nm},P)| \leq  1, \text{ for all } \ell=0,1,2,3\dots, I=0,1,2\,,
\end{equation}
\item FESR sum rules
\be
\begin{split}
&(\text{mean}-\e \times \text{error})<F_n<(\text{mean}+\e\times\text{error}),~n=0,1,2\,.
\end{split}
\ee
\end{enumerate}

Convergence with $N, L, P$ are shown in figure \eqref{fig:LPconvergence} for the weakest choice of tolerance $\epsilon=1$. Truncating the spin at $L=9$ and $P=10$ does not alter the third significant digits. Hence, throughout our analysis, we use these truncations. The convergence with $N$ is evident in figure \eqref{fig:LPconvergence}. For light quark contribution, the final bound in the second step is $680.0^{+3}_{-3}\times 10^{-10}$, which shows improvement from full unitarity. This is weakest possible lower bound becasue of weakest choice of tolerance.  We want to draw attention to the fact that this is a significant improvement over the weakest lower bound found in $a_\mu^\text{LO-HVP}> 623\times 10^{-10}$ found in \cite{qcdpaper}. Note that lower bound in \cite{qcdpaper} $a_\mu^\text{LO-HVP}> 657_{-34}^{+34}\times 10^{-10}$ with errors are due to FESRs so the weakest possible bound in \cite{qcdpaper} $a_\mu^\text{LO-HVP}> 623\times 10^{-10}$. 

\textit{\textcolor{Blue}{Step 3$\setminus$}  Imposing chiral symmetry breaking:}
The theory of pion well approximates the low energy QCD due to chiral symmetry breaking. We use tree level $\chi$PT to capture the low energy physics. These barely improve the bound (adds half to the third significant digit), but we impose these for completeness.

\subsection{Outcome}
The minimization process will optimizes the parameters $d_n, c_n, a_{nm}, b_{nm}$ until it satisfy all the unitary conditions and the FESRs then return the values of $d_n, c_n, a_{nm}, b_{nm}$ that minimised $a_\mu^\text{LO-HVP}[d_n,N]$. Note that the set of $d_n$ in these process that will provide Im$\Pi(s)$. As a final outcome SDPB solver will return Min$a_\mu^\text{LO-HVP}$ and the all parameters, particularly $d_n$. Using these values, we can construct $\rho_1^1(s)$ via Eq.~\eqref{ansatzrhoF}. Using this construction, we plot the spectral density in fig \eqref{fig:expdata} in main text. With the obtained $d_n$, we reconstruct $\rho_1^1(s)$ and use it to plot the spectral density. The resulting plot, shown in fig \eqref{fig:expdata} of the main text, illustrates the behaviour of the spectral function which determine Min$a_\mu^\text{LO-HVP}$.


\section{Different observables and the bootstrapped spectral density}

\subsection{{Window observables}}
{
The formula for the HVP contribution as given in main text
\begin{equation}\label{hvp_lo_win}
    a_\mu^\text{LO-HVP} = \frac{4\alpha^2}{\pi} \int_{4m_\pi^2}^{\infty}\frac{K(s) \text{Im}\Pi(s)}{s} \, ds\,,
\end{equation}
The lattice QCD computation mostly uses the time-momentum representation \cite{RBC}
\begin{equation}
a_\mu^{\mathrm{HVP}} = \left( \frac{\alpha}{\pi} \right)^2 \int_0^\infty \mathrm{d}t \, \tilde{K}(t) G(t),
\end{equation}
Windows Observables in Euclidean time are defined by 
\begin{small}
\begin{align}
\Theta_{\mathrm{SD}}(t) &= 1 - \Theta(t, t_0, \Delta), \quad 
\Theta_{\mathrm{win}}(t) = \Theta(t, t_0, \Delta) - \Theta(t, t_1, \Delta), \nonumber \\
\Theta_{\mathrm{LD}}(t) &= \Theta(t, t_1, \Delta), \quad 
\Theta(t, t', \Delta) = \frac{1}{2} \left( 1 + \tanh\frac{t - t'}{\Delta} \right),
\end{align} 
\end{small}
as an additional weight function with parameters
\begin{equation}
t_0 = 0.4\,\mathrm{fm}, \quad t_1 = 1.0\,\mathrm{fm}, \quad \Delta = 0.15\,\mathrm{fm}.
\end{equation}
The weight functions for \eqref{hvp_lo_win}\cite{window}
\begin{align}
\tilde{\Theta}(s) &= \frac{3s^{5/2}}{8m_\mu^4} \hat{K}(s) \int_0^\infty \mathrm{d}t \, \Theta(t) e^{-t\sqrt{s}} 
\int_0^\infty \mathrm{d}s' \, w\left( \frac{s'}{m_\mu^2} \right) \nonumber \\
&\quad \times \left( t^2 - \frac{4}{s'} \sin^2 \left( \frac{t\sqrt{s'}}{2} \right) \right),\\
&w(r) = \frac{\left[r + 2 - \sqrt{r(r + 4)} \right]^2}{\sqrt{r(r + 4)}}, ~\hat{K}(s) =\frac{3 s}{m_{\mu }^2}K(s).
\end{align}
Window observables for HVP from data-driven approach \cite{window}, from lattice QCD and phenomenology are shown in \eqref{tab:window} along with bootstrapped spectral density for comparison in units of $10^{-10}$. For bootstrap spectral density we took the case when $\e=1$ and $N=95$ (the one presented in figure \eqref{fig:expdata}) for which the lower bound is $682\times 10^{-10}$ (\textbf{conservative lower bound}) and is consistent with SM --see table \eqref{tab:e}. The intermediate window contribution from bootstrap is quite big compared to others, while SD, LD contributions are small, because bootstrapped spectral density has only one peak near rho-resonance, rest almost zero.}
\begin{table}
\begin{footnotesize}
\begin{tabular}{lllll}
\toprule
 & $a_\mu^{\mathrm{HVP}}{}_{\mathrm{SD}}$ & $a_\mu^{\mathrm{HVP}}{}_{\mathrm{int}}$ & $a_\mu^{\mathrm{HVP}}{}_{\mathrm{LD}}$ & $a_\mu^{\mathrm{HVP}}{}_{\mathrm{total}}$ \\
\midrule
Data-driven \cite{window} & 68.4(5) & 229.4(1.4) & 395.1(2.4) & 693.0(3.9) \\
\midrule
RBC/UKQCD \cite{RBC} & – & 231.9(1.5) & – & 715.4(18.7) \\
BMWc \cite{lat1} & – & 236.7(1.4) & – & 707.5(5.5) \\
BMWc/KNT \cite{Mz,lat1} & – & 229.7(1.3) & – & – \\
Mainz/CLS \cite{Ce:2022kxy} & – & 237.30(1.46) & – & – \\
ETMC \cite{ref100} & 69.33(29) & 235.0(1.1) & – & – \\
\midrule
\textbf{Bootstrap} & \textbf{63} & \textbf{335} & \textbf{284} & \textbf{682} \\
\bottomrule
\end{tabular}
\end{footnotesize}
\caption{{Comparison: Window observables for SD, intermediate , LD contributions. The intermediate from bootstrap is quite big than others, while SD, LD contributions are small, because bootstrapped spectral density has only one peak near rho-resonance, rest almost zero.}}
\label{tab:window}
\end{table}
\subsection{{Two pion contribution}}
\begin{figure}[hbt!]
     \centering
         \includegraphics[scale=0.4]{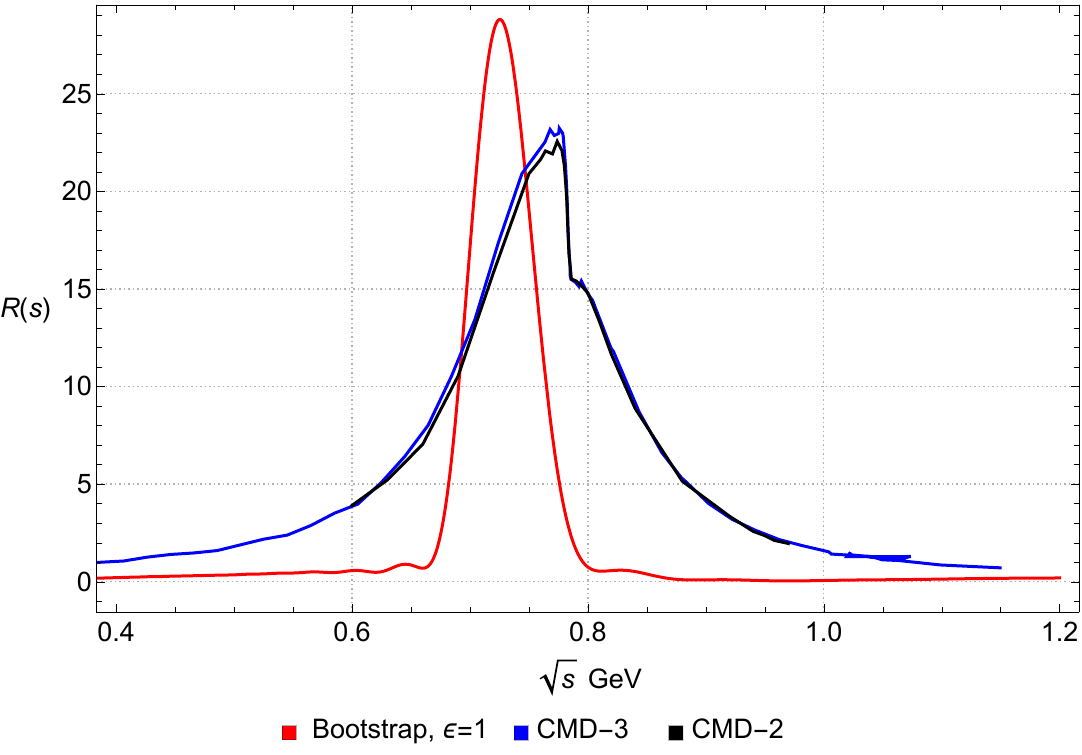}
        \caption{{Comparison of bootstrap spectral density with the measured CMD-3 and CMD-2 pion form factor (properly normalised).}}
        \label{fig:cmd}
\end{figure}

Since, the discrepancy between the theory and experiment lies mainly in the
two-pion channel. We can compare our bootstrap spectral density with CMD pion-pion form factor data\cite{CMD-3,CMD-2} in the energy range $\sqrt{s}=0.327$ to $1.2$ GeV. Note that for comparison with $R(s)=12\pi \text{Im}\Pi(s)$ we have to multiply an extra normalization factor (see normalization in eq \ref{semimat}) namely, we compare with $12\pi\times\left(\frac{\sqrt{\frac{4 \pi }{3}} \left(\frac{s-4}{4}\right)^{3/4}}{\left(8 \pi ^3\right) \sqrt[4]{s}}F_\pi(s)\right)^2$, and we take $F_\pi^2(s)$ from CMD-3 and CMD-2 pion-pion form factor data--see figure \eqref{fig:cmd}. In the energy range $\sqrt{s}=0.327$ to $1.2$ GeV, from bootstrap we have $$a_\mu^{had-bootstrap}(\sqrt{s}=0.327-1.2)= 664\times 10^{-10}\,,$$ while CDM-3 reported the $$a_\mu^{had}(2\pi,\text{CMD-3})= 526(4.2)\times 10^{-10}\,,$$ and in \cite{th1} $$a_\mu^{had}(2\pi,[3])= 506(3.8)\times 10^{-10}.$$ Note that $a_\mu^{had-bootstrap}(\sqrt{s}=0-0.327)= 4.8\times 10^{-10}$ and $a_\mu^{had-bootstrap}(\sqrt{s}=1.2-\infty)= 13\times 10^{-10}$ giving total $682\times 10^{-10}$ for $N=95,\e=1$. Note that with the two pion contribution adding the rest of the contributions in [3] it was reported $$a_\mu^{had-LO}([3])=693(3.9)\times 10^{-10}$$ and \cite{CMD-3} reported (just removing the $2\pi$ contribution and adding CMD-3 result) $$a_\mu^{had-LO}(CMD-3)=714(4.2)\times 10^{-10},$$ both \textbf{respect our  conservative lower bound} (adding with charmonium and bottomonium resonance contributions \cite{charm1}) $$\text{Min}_{bootstrap}[a_\mu^\text{had-LO}]=
688.4^{+3}_{-3}\times 10^{-10}\,.
$$ Further note that bootstrap contribution for $\sqrt{s}=0.327$ to $1.2$ is bigger than others, because bootstrapped spectral density has only one peak near rho-resonance, rest almost zero.

\section{QCD finite energy sum rules}\label{ap:qcdsumrules}
For convenience, we introduced the following notation for the FESRs
$$
F_n\equiv\frac{1}{s_0^{1+n}}\int_{4}^{s_0}t^n  \frac{\mathrm{Im}\Pi(t)}{(2\pi)^4}dt, ~n=0,1,2\,,
$$
where we have omitted the explicit $s_0$ dependence, as it will be fixed below. We use the following QCD sum rules \cite{qcdpaper}
\be
\begin{split}
F_0&=\frac{\pi}{s_0}\frac{1}{(2\pi)^4}\left(\frac{4}{9} F_0^{(up)}(s_0)+\frac{1}{9} F_0^{(down)}(s_0)+\frac{1}{9} F_0^{(strange)}(s_0)\right)\,,\\
F_1&=\frac{\pi}{s_0^2}\frac{1}{(2\pi)^4} \left(\frac{4}{9} F_1^{(up)}(s_0)+\frac{1}{9} F_1^{(down)}(s_0)+\frac{1}{9} F_1^{(strange)}(s_0)\right)\,,\\
F_2&=\frac{\pi}{s_0^3}\frac{1}{(2\pi)^4} \left(\frac{4}{9} F_2^{(up)}(s_0)+\frac{1}{9} F_2^{(down)}(s_0)+\frac{1}{9} F_2^{(strange)}(s_0)\right)\,,
\label{eq:uds_sums}
\end{split}
\ee
with
\begin{widetext}
\begin{align}
F_0^{(q)}(s_0) &= \frac{1}{4 \pi^2} \Big[ 1 + \frac{\alpha_s(\mu)}{\pi} T_{10} + \left( \frac{\alpha_s(\mu)}{\pi} \right)^2 (T_{20} + T_{21}) + \left( \frac{\alpha_s(\mu)}{\pi} \right)^3 (T_{30} + 2 T_{31} + 2 T_{32}) \\
&+ \left( \frac{\alpha_s(\mu)}{\pi} \right)^4 (T_{40} + 2 T_{41} + 6 T_{42} + 6 T_{43}) \Big] s_0 - \frac{3}{2 \pi^2} m_q^2, \\
F_1^{(q)}(s_0) &= \frac{1}{8 \pi^2} \Big[ 1 + \frac{\alpha_s(\mu)}{\pi} T_{10} + \left( \frac{\alpha_s(\mu)}{\pi} \right)^2 (T_{20} + T_{21}) + \left( \frac{\alpha_s(\mu)}{\pi} \right)^3 (T_{30} + T_{31} + T_{32}) \\
&+ \left( \frac{\alpha_s(\mu)}{\pi} \right)^4 \left( T_{40} + \frac{1}{3} T_{41} + \frac{2}{3} T_{42} + \frac{3}{4} T_{43} \right) \Big] s_0^2 - 2 m_{q}\langle\bar{q} q\rangle \left( 1 + \frac{\alpha_s(\mu)}{3}  \right) - \frac{1}{12 \pi} \langle \a_s G^2\rangle \left( 1 + \frac{7}{6}\frac{\alpha_s(\mu)}{\pi}  \right) , \\
F_2^{(q)}(s_0)&= \frac{1}{12 \pi^2} \Big[ 1 + \frac{\alpha_s(\mu)}{\pi} T_{10} + \left( \frac{\alpha_s(\mu)}{\pi} \right)^2 \left( T_{20} + \frac{1}{3} T_{21} \right) + \left( \frac{\alpha_s(\mu)}{\pi} \right)^3 \left( T_{30} + \frac{1}{3} T_{31} + \frac{2}{9} T_{32} \right)  \\
&\quad +  \left( \frac{\alpha_s(\mu)}{\pi} \right)^4 \left( T_{40} + \frac{1}{3} T_{41} + \frac{2}{9} T_{42} + \frac{2}{9} T_{43} \right) \Big] s_0^3 - \frac{224}{81} \pi \alpha_s(\mu) m_q \langle \bar{q} \bar{q} q q\rangle\,,
\end{align}
\end{widetext}
where $\mu=\sqrt{s_0}$ and
\be 
F_k^{(q)}(s_0)=\int_{4m_\pi^2}^{s_0} \frac{\text{Im}\Pi_q(t)}{\pi} t^k dt
\ee
We numerically solve four loops RG equation for $\alpha_s(\mu)$ using $\alpha_s (M_\tau)$ as a boundary condition, which is above the charm threshold. Since we are interested in up to $s_0=1.19$ GeV${}^2$, below the charm threshold, we start the RG for $N_f=4$, do matching at the charm threshold, and then transit to $N_f=3$. The RG equation is as follows \cite{fourloopbeta}
\be
\frac{1}{2} \, \mu \, \frac{\partial a(\mu)}{\partial \mu} = -\sum_{n=0}^3 a(\mu)^{n+2} \, \beta_n\,,
\ee
where $\alpha_s(\mu)=4\pi a(\mu)$ with 
\begin{align*}
\beta_0 &= 11 - \frac{2}{3} N_f, ~~
\beta_1 = 102 - \frac{38}{3} N_f, \\
\beta_2 &= 1428.5 - 279.611 \, N_f + 6.01852 \, N_f^2, \\
\beta_3 &= 29243 - 6946.3 \, N_f + 405.089 \, N_f^2 + 1.49931 \, N_f^3.
\end{align*}
Leading order RG effect of quark mass is also taken care of \cite{quarkmass}. RG effect for the condensates also has been take care of up to NLO using the fact that $m_q\langle \bar{q}q\rangle$ and $\langle\beta  G^2\rangle +4\gamma m_q\langle \bar{q}q\rangle$ does not run with RG,  where $\beta$ is the beta function $\beta(\mu)=\mu  \frac{\partial \alpha _s(\mu )}{\partial \mu }$ and $\gamma$ is mass anomalous dimension $ \mu  \frac{\partial m_q(\mu )}{\partial \mu }=-\gamma(\mu) m_q(\mu)$.
\\
We used QCD parameters as given below in tables \eqref{tabqcd1}, \eqref{tabqcd2}--see \cite{qcdpaper} and references \cite{fourloopbeta,PDG2022, quarkmass,refpara, refpara_k, AlamKhan:2023dms}, 
\begin{table}[t]
\centering
\begin{tabular}{|c|c|}
\hline
\textbf{Coefficient} & \textbf{Value} \\
\hline
\( T_{10} \) & \( 1 \) \\
\( T_{20} \) & \( 1.63982 \) \\
\( T_{21} \) & \( \frac{9}{4} \) \\
\( T_{30} \) & \( -10.2839 \) \\
\( T_{31} \) & \( 11.3792 \) \\
\( T_{32} \) & \( \frac{81}{16} \) \\
\( T_{40} \) & \( -106.896 \) \\
\( T_{41} \) & \( -46.2379 \) \\
\( T_{42} \) & \( 47.4048 \) \\
\( T_{43} \) & \( \frac{729}{64} \) \\
\hline
\end{tabular}
\caption{Coefficients for \( N_f = 3 \).}
\label{tabqcd1}
\end{table}
\begin{table}[t!]
\centering
\begin{tabular}{|c|c|}
\hline
\textbf{Parameter} & \textbf{Value} \\
\hline
$\alpha$ & $1/137.036$ \\
$\alpha_s(M_\tau)$ & $0.312 \pm 0.015$ \\
$m_u(2 \, \mathrm{GeV})$ & $2.16^{+0.49}_{-0.26} \, \mathrm{MeV}$ \\
$m_d(2 \, \mathrm{GeV})$ & $4.67^{+0.48}_{-0.17} \, \mathrm{MeV}$ \\
$m_s(2 \, \mathrm{GeV})$ & $0.0934^{+0.0086}_{-0.0034} \, \mathrm{GeV}$ \\
$f_\pi$ & $(0.13056 \pm 0.00019)/\sqrt{2} \, \mathrm{GeV}$ \\
$m_n\langle \bar{n}n \rangle$ & $-\frac{1}{2} f_\pi^2 m_\pi^2$ \\
$m_s\langle \bar{s}s \rangle$ & $r_m r_c m_n \langle \bar{n}n \rangle$ \\
$r_c$ & $0.66 \pm 0.10$ \\
$m_s/m_n = r_m$ & $27.33^{+0.67}_{-0.77}$ \\
$\langle \alpha G^2 \rangle$ (2 GeV) & $0.0649 \pm 0.0035 \, \mathrm{GeV}^4$ \\
$\kappa$ & $3.22 \pm 0.5$ \\
$\kappa \alpha_s \langle \bar{n}n \rangle^2$ & $\kappa (1.8 \times 10^{-4}) \, \mathrm{GeV}^6$ \\
$\alpha_s \langle (\bar{s}s)^2 \rangle$ & $r_c^2 \alpha_s \langle \bar{n}n \rangle^2$ \\
\hline
\end{tabular}
\caption{QCD parameters and values with $m_n=(m_u+m_d)/2, \langle \bar{n}n \rangle=\langle \bar{u}u \rangle=\langle \bar{d}d \rangle $.}
\label{tabqcd2}
\end{table}

Using Holder's inequality and positivity of Im$\Pi_q(t)$,  the paper \cite{qcdpaper} established that each quark sector should obey the following inequality 
$$
\left(\frac{F_1^{(q)}}{(4m_\pi^2)^2} - F_B\right)^2 \leq \left(\frac{F_1^{(q)}}{(4m_\pi^2)^2} - (F_0^{(q)})^2/F_1^{(q)}\right)^2,
$$
with $F_B=\frac{F_0^{(q)}}{4m_\pi^2}-\frac{\left(\frac{F_1^{(q)}}{4m_\pi^2}-F_0^{(q)}\right)^2}{\frac{F_2^{(q)}}{4m_\pi^2}-F_1^{(q)}}$, we have suppressed the $s_0$ labels from $F_k^{(q)}(s_0)$ for clarity and $q$=up, down, strange. One can easily verify that for up and down quark, it gets violated below $s_0=1.09$ GeV${}^2$ while for strange quark, it gets violated below $s_0=1.19$ GeV${}^2$--see \cite{qcdpaper}. 

We will be slightly conservative about the choice of $s_0$, namely, we will choose $s_0=1.19$ GeV${}^2$ uniformly for all three of $u,d,s$. While it is, in principle, possible to lower this value, as the combined contributions from u,d,s in \eqref{eq:uds_sums} do not immediately violate the inequality mentioned above, just immediately below $s_0=1.19$ GeV${}^2$. One should be cautious about going too small, $s_0$. At sufficiently low $s_0$, the theoretical control over the Operator Product Expansion (OPE) begins to weaken. Specifically, the OPE may break down, or its truncation error may become significant--see enlightening discussion in \cite{SVZ} page 397. Ideally, one would include a quantitative estimate of these kinds of uncertainties to ensure the robustness of results at low $s_0$--we leave this as future exploration.

\end{document}